\documentclass[12pt,notoc]{JHEP3}

\usepackage{amsmath,amssymb,euscript,array,cite,mathrsfs,amsfonts}
\usepackage{epsfig}

\newcommand*{\Dsl}[0]{{\rlap{\kern2.25pt /}{D}}}
\newcommand*{\Asl}[0]{{\rlap{\kern2.25pt /}{A}}}
\newcommand*{\dsl}[0]{{\rlap{\kern0.5pt /}{\partial}}}
\newcommand*{\xisl}[0]{{\rlap{\kern0.5pt /}{\xi}}}
\newcommand*{\asl}[0]{{\rlap{\kern0.5pt /}{a}}}
\newcommand*{\bsl}[0]{{\rlap{\kern0.5pt /}{b}}}

\def\Dslash{\,\,{\raise.15ex\hbox{/}\mkern-12mu D}}

\newcommand{\SP}[1]{\begin{equation}\begin{split} #1
\end{split}\end{equation}}

\newcommand{\Tr}{\operatorname{Tr}}

\def\B0{{\boldsymbol 0}}

\def\Tr{{\rm Tr}}

\def\Dbarslash{\,\,{\raise.15ex\hbox{/}\mkern-12mu {\bar D}}}
\def\Dslash{\,\,{\raise.15ex\hbox{/}\mkern-12mu D}}
\def\delslash{\,\,{\raise.15ex\hbox{/}\mkern-9mu \partial}}
\def\delbarslash{\,\,{\raise.15ex\hbox{/}\mkern-9mu {\bar\partial}}}

\def\PP{\mathscr{P}}

\newcommand{\EQ}[1]{\begin{equation}\begin{split} #1
\end{split}\end{equation}}

\title{Numerical Study of the Two Color Attoworld}
\author{Simon Hands, Timothy J. Hollowood and Joyce C. Myers\\ 
  Physics Department, Swansea University,\\ Singleton Park, Swansea SA2 8PP, UK\\ E-mail:
  \email{s.hands@swan.ac.uk, t.hollowood@swansea.ac.uk, j.c.myers@swan.ac.uk}}

\abstract{We consider QCD at very low temperatures and non-zero quark chemical potential 
from lattice Monte Carlo simulations of the two-color theory in a very small spatial volume
(the attoscale). In this regime the quark number rises in discrete levels in
qualitative agreement with what is 
found analytically at one loop on $S^3 \times S^1$ with
$R_{S^3}\ll\Lambda_{QCD}^{-1}$ \cite{Hands:2010zp}. The detailed level degeneracy,
however, cannot be accounted for using weak coupling arguments. At each rise in the quark number
there is a corresponding spike in the Polyakov line, also in agreement with the
perturbative results. In addition the quark number susceptibility shows
a similar behaviour to the Polyakov line and appears to be a good
indicator of a confinement-deconfinement type of transition.}


\preprint{}

\begin{document}


\section{Introduction}

The study of QCD at low temperature and non-zero chemical potential $\mu$ is
complicated by the presence of the sign problem: setting $\mu\not=0$
results in a complex action which prevents the importance sampling
necessary for conventional methods of lattice simulation. Perturbation theory is
also not valid in this regime in the infinite volume limit because the coupling
strength is large. One method of avoiding the sign problem is to study a theory
which may closely resemble QCD, specifically the two-color theory, for which the
action is real even for non-zero chemical potential. This is the theory we study
in this paper. Another technique which works in the low temperature and non-zero
chemical potential regime is to compactify the spatial volume onto a manifold
with size $R \ll \Lambda_{QCD}^{-1}$; this is the limit of the attoworld and here it is
possible to calculate using perturbation theory. 

In \cite{Hands:2010zp} we calculated the phase diagram of QCD in the $\mu R$-$T
R$ plane on $S^3 \times S^1$, where $R$ is the radius of $S^3$, considering both
$N_c = 3$, and also $N_c = \infty$ for which there is a thermodynamic limit and
genuine phase transitions can occur.  In a small volume, quarks occupy
well-defined single-particle states of finite degeneracy, whose energies are
well-separated on the scale defined by $T$ (given by the inverse radius of the
$S^1$).  In the large $N_c$ theory, we found that for $RT\ll1$ there is a rich
structure of third-order Gross-Witten-like transitions between ``confined" and
``deconfined" phases (the names refer to the behaviour of the Polyakov line) as
the chemical potential $\mu$ passes one of the quark energy levels. The
interpretation is that as $\mu$ scans past a level, the quarks fill the level
and the system becomes de-confined in the sense that the Polyakov line gains a
non-vanishing expectation value. 
As $\mu$ increases further, the system re-enters the confined phase where
the Polyakov line vanishes. The Polyakov line consequently exhibits a peak
with discontinuous slope where the phase transitions occur. At finite $N_c$, similar
peaks are seen, but in this case the curve is smooth since there are no phase
transitions away from the thermodynamic limit. 

The analysis of  \cite{Hands:2010zp} was
done in the one-loop approximation and it is important to consider
non-perturbative effects.  In this paper we calculate several observables using
lattice gauge theory simulations of QCD with $N_c = 2$ (also known as Two Color
QCD or QC$_2$D) formulated on small tori, and
at very low temperatures, using $L_s^3\times L_t=3^3 \times 64$ lattices. We
will find that the quark number, Polyakov line, and quark number susceptibility
qualitatively resemble the results from perturbation theory on $S^3 \times S^1$
\cite{Hands:2010zp} but with noticeable quantitative differences, which may
result from working at larger coupling strength, or may be due to the
formulation of the theory on a different manifold, an effect which may be
possible in the small volume limit in which we work. We supplement further the
perturbative results of \cite{Hands:2010zp} with numerical calculations of the
relevant observables on $S^3 \times S^1$ for $N_c = 2$. In neither case is 
there a  thermodynamic limit, as both $N_c$ and $R$ ($L_s$) are finite.
These findings reinforce our observation, made in ~\cite{Hands:2010zp},
that deconfining behaviour in gauge theories with $\mu\not=0$
appears to be associated with a non-zero density of
gapless quark states.

In the next section we introduce the lattice formulation of QCD with gauge group
SU(2), and in Section~\ref{sec:free} briefly outline the behaviour of the lattice
model on a finite system as $\mu$ is increased in the non-interacting limit. An
important difference with the perturbative approach of \cite{Hands:2010zp} is
that as well as an IR cutoff $L_s\sim R$, there is in this case an explicit UV
scale associated with the lattice spacing $a$. Section~\ref{sec:results}
presents results from non-perturbative lattice simulations of the interacting
theory, where possible comparing them with the equivalent quantities 
calculated on $S^3\times S^1$ with $N_c=2$ and $N_c=\infty$. We conclude with a
brief discussion.

\section{Formulation and Simulation of the Lattice Model}

We begin by defining the action for Two Color QCD on a hypercubic
lattice~\cite{Hands:2006ve}, choosing units where the lattice spacing $a=1$, and
a Euclidean spacetime index $\nu=0,\ldots,3$:
\begin{equation}
S=\sum_{x,y,\alpha} \bar\psi_x^\alpha M_{xy}[U;\mu]\psi_y^\alpha-{\beta\over
N_c}\sum_{x,\nu<\lambda}\mbox{tr}U_{\nu\lambda x},
\end{equation}
where $N_c=2$, $U_{\nu\lambda}$ is the oriented product of 4
SU(2)-valued link fields $U_{\nu x}$ around the sides of an elementary plaquette
in the $\nu$-$\lambda$ plane, 
$\psi$, $\bar\psi$ are Grassmann-valued quark fields located on the lattice
sites, whose index $\alpha$ runs over $N_f=2$ flavors, and $\mu$ is the quark
chemical potential. The quark matrix utilises
the Wilson formulation for lattice fermions:
\begin{equation}
M_{xy}[U;\mu]=\delta_{xy}-\kappa\sum_\nu\biggl[(1-\gamma_\nu)e^{\mu\delta_{\nu
0}}U_{\nu x}\delta_{y,x+\hat\nu}+(1+\gamma_\nu)e^{-\mu\delta_{\nu 0}}U_{\nu
y}^\dagger\delta_{y,x-\hat\nu}\biggr].
\label{eq:M}
\end{equation}
The parameter $\beta\equiv2N_c/g^2$, where $g$ is the bare Yang-Mills coupling,
and the hopping parameter $\kappa$ is
related to the bare quark mass $m$ via 
\begin{equation}
m={1\over{2\kappa}}-{1\over{2\kappa_c(\beta)}}.
\end{equation}
In the free field limit $\beta\to\infty$
$\kappa_c={1\over8}$, but since the
action defined via (\ref{eq:M}) has no manifest chiral symmetry, its value 
is subject to quantum corrections and must in general be determined
by simulation; chiral symmetry is then only recovered in the limit
$\kappa\to\kappa_c$. Chemical potential is introduced via the orthodox
prescription of treating $\mu$ as a constant imaginary timelike abelian gauge potential
\cite{mu}.

For the system in hand it is possible to calculate quantum corrections non-perturbatively using 
numerical Monte Carlo simulation~\cite{Hands:2006ve,Hands:2010gd}. The
simulation proceeds via an orthodox hybrid Monte Carlo (HMC) algorithm, which
unlike the case of QCD is not subject to the notorious Sign Problem for
$\mu\not=0$, since the functional measure $\mbox{det}^{N_f}M$ remains real and
therefore positive due to the special SU(2) property
\begin{equation}
KM(\mu)K^{-1}=M^*(\mu)\;\;\;\mbox{with}\;\;\;K\equiv C\gamma_5\tau_2,
\end{equation}
where the Pauli matrix $\tau_2$ acts on color indices.
The most important thermodynamic
observable in the presence of chemical potential is the quark density
\begin{equation}
n_q={T\over V}{{\partial\ln{\cal Z}}\over{\partial\mu}}=\sum_\alpha\kappa
\biggl\langle\bar\psi^\alpha_x(\gamma_0-1)e^\mu U_{0x}\psi^\alpha_{x+\hat0}
+\bar\psi^\alpha_x(\gamma_0+1)e^{-\mu} U_{0x-\hat0}^\dagger\psi^\alpha_{x-\hat0}
\biggr\rangle.
\label{quark_dens}
\end{equation}
The other key observable in this study is the Polyakov line, defined in terms
of link variables by
\begin{equation} {\mathscr P}={1\over L_s^3}\sum_{\vec x}{1\over
N_c}\mbox{tr}\biggl\langle\prod_{t=1}^{L_t}U_{0\vec x,t}\biggr\rangle .
\label{eq:L} 
\end{equation}

\section{Free field limit results ($\beta = \infty$)}
\label{sec:free}

\begin{figure}[t]
\center
\includegraphics[width=13cm]{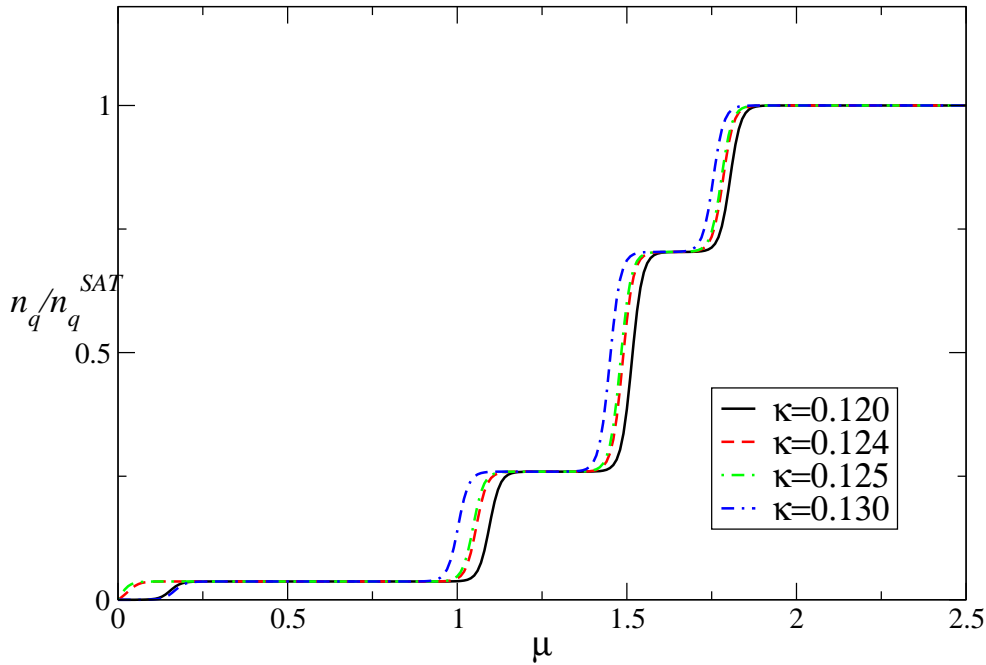}
\caption{Expectation value of quark density as a function of $\mu$
for free quarks on a $3^3\times64$ lattice.}
\label{fig:free}
\end{figure}

In the free field limit $U_\nu=1$ the quark density (\ref{quark_dens}) on a finite lattice can be
evaluated via a simple mode sum~\cite{Hands:2006ve}. The results for a
$3^3\times64$ system are shown as a function of $\mu$ for various $\kappa$ in
Fig.~\ref{fig:free}.
The numbers are expressed in the form $n_q/n_q^{\rm SAT}$, 
where $n_q^{\rm SAT}=2N_cN_f$ is the maximum quark density which can be
accommodated on the lattice as a result of the Pauli principle, essentially one
quark of each spin, flavor and color per site. The existence of such a
``saturated'' system is a consequence of having an explicit ultra-violet cutoff;
there is no corresponding feature in the results of \cite{Hands:2010zp} which were obtained
in the continuum.

Fig.~\ref{fig:free} shows $n_q$ increasing monotonically with $\mu$,
but with a step-wise behaviour rather than the $n_q\propto\mu^3$
expected in the thermodynamic zero-temperature limit. A physical way of understanding
this is that for a system with $T^{-1}=L_t\gg L_s$, the thermal smearing of the
Fermi surface associated with
degenerate quarks will be much smaller than
the $O(L_s^{-1})$ spacing between plane wave modes, meaning that the
Fermi surface will be markedly non-spherical~\cite{Hands:2002mr}.
Indeed, close inspection of the figure reveals that the total quark occupancy 
${\mathscr N}=2N_cN_f{\cal G}$ on the
plateaux have  respectively ${\cal G}=1$, 7, 19 and 27, implying the existence of
well-separated energy levels (on the scale of $T$) with degeneracies 1, 6, 12 and
8. These are readily identified with the plane-wave states available on a
$3^3$ system, which in units of the mode spacing $2\pi/L_s$ take the form
$(0,0,0)$, $(\pm1,0,0)$, $(\pm1,\pm1,0)$ and $(\pm1,\pm1,\pm1)$ respectively.
The figure also shows that varying $\kappa$ has a small effect on the energies
of these discrete levels, given by the value of $\mu$ at the riser of each step,
but none on their degeneracy. For the value $\kappa=0.125$ corresponding to free
massless quarks, the onset at which  $n_q$ rises from zero takes place at
$\mu=0$; this threshold shifts to larger $\mu$ for $\kappa\not=\kappa_c$.

The results for $\beta=\infty$ are thus in qualitative accord with the one-loop findings
of \cite{Hands:2010zp}, except in that case the equivalent degeneracy of the $L$th level is given
by ${\mathscr N}_L=2 N_c N_f \sum_{l=1}^{L} \ell(\ell+1)$, 
with $\ell=1,2,\ldots$ as appropriate for angular momentum
eigenstates on a hypersphere. Another difference is that for massless quarks the
hypersphere levels $\varepsilon_\ell=(\ell+{1\over2})R^{-1}$ are equally spaced.
It is thus reasonable to anticipate that the
structure of Fig.~\ref{fig:free} would remain stable under weakly coupled quantum 
corrections, {\it i.e.\/}~with $\beta$ large but finite.

\section{Results for the Interacting Case}
\label{sec:results}

In this section, we first present numerical results obtained with $\kappa=0.124$ at two
coupling strengths $\beta=24$ and $\beta=6$, close to the weak-coupling limit.
This will enable an estimate of how quantum corrections evolve with coupling,
but it should be noted that both values are considerably weaker than those
employed in the studies ~\cite{Hands:2006ve,Hands:2010gd} which explored
significantly larger volumes. The lattice spacing at $\beta=1.9$ was determined
via the string tension to be $a\simeq0.19$fm, implying that $L_s=3$ corresponds
to a system size 
considerably smaller than a fermi even at these stronger couplings. Using the
one-loop beta-function estimate
$a^\prime/a=\exp[-(\beta^\prime-\beta)/4b_1N_c]$ with
$b_1=(11N_c-2N_f)/48\pi^2$, we can be confident that the simulations
described here definitely probe the attoworld. Another important difference is
that in previous work a diquark source of the form
$j\psi\psi+\bar\jmath\bar\psi\bar\psi$ (in effect a Majorana mass for the
quarks) was introduced; this had the effect of mitigating infra-red fluctuations
associated with the Goldstone modes induced by superfluidity due to diquark
condensation, thus improving the
performance of the HMC algorithm. In the current work we judged this would not
be needed for such small systems; even so it was found that, eg. the number of
conjugate gradient iterations required for the HMC acceptance step occasionally
exceeded $10^3$ (see Fig~\ref{fig:congrad} below). 


\begin{figure}[htbp]
\centering
\includegraphics[width=11.5cm]{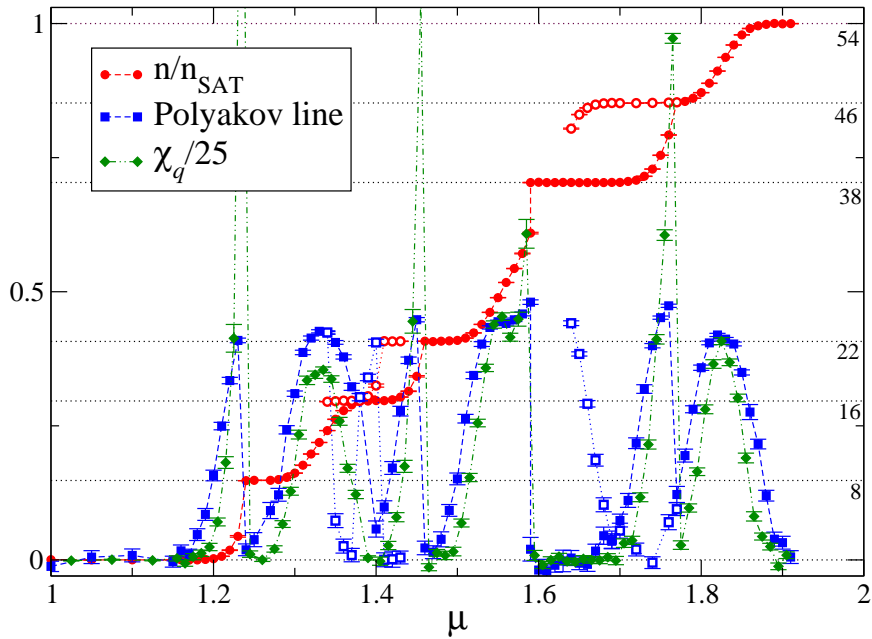}
\caption{Plot of $n_q/n_q^{\rm SAT}$, ${\mathscr P}$ and quark number susceptibility $\chi_q$ (rescaled
by a factor 1/25) versus
$\mu$ for $\beta=24$, $\kappa=0.124$ on $3^3\times64$. Numbers on the right hand
side should be multiplied by $2N_f$ to give the occupancy ${\mathscr N}$.
The meaning of the open symbols is discussed in the text}
\label{fig:muscan}
\end{figure}
\begin{figure}[htbp]
\vspace{8mm}
\centering
\includegraphics[width=9.5cm]{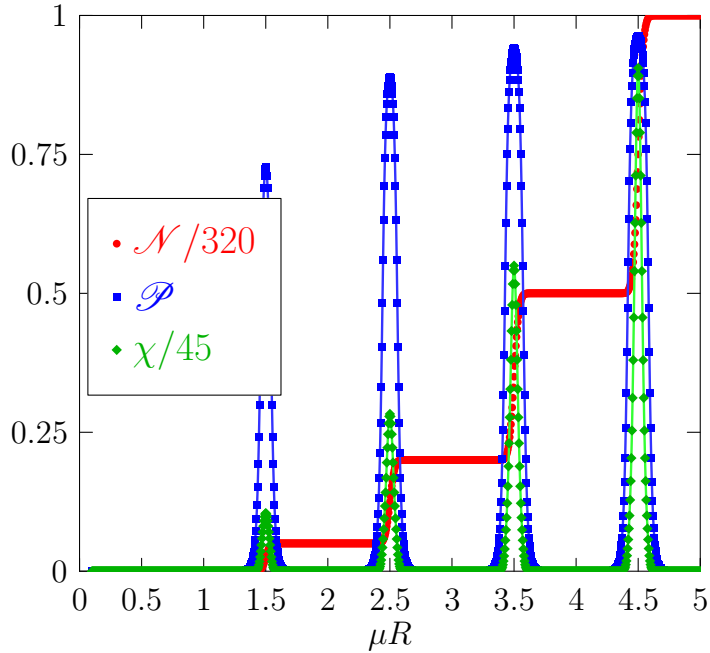}
\caption{Plot of quark number ${\mathscr N}$, Polyakov line ${\mathscr P}$ and quark number susceptibility $\chi$ for $N_c = 2$ on $S^3 \times S^1$. $N_f = 2$, $m R = 0$, $\frac{1}{T R} = 50$.}
\label{all_sphere}
\end{figure}

Fig.~\ref{fig:muscan} shows the results of a scan along the $\mu$-axis for
$\beta=24$. For each value of $\mu$ 2000 HMC trajectories of
mean length 0.5 were accumulated. The quark density $n_q$ is calculated
according to eq. (\ref{quark_dens}), and we define the quark number susceptibility by
its derivative with respect to $\mu$, 
\begin{equation} \chi_q\equiv{{\partial
n_q}\over{\partial\mu}}. 
\label{eq:susc}
\end{equation}
In principle in lattice QCD $\chi_q$ is calculated
in terms of correlations of quark bilinears~\cite{Gottlieb:1987ac}, but in this
study since we have a finely-spaced comb of $\mu$-values we adopt the more
pragmatic route of approximating the derivative in (\ref{eq:susc}) by a finite
difference.

It is interesting to compare the lattice results with those of a
one-loop calculation on $S^3 \times S^1$. Figure \ref{all_sphere} shows the same
quantities (though normalized differently) on $S^3 \times S^1$ and plotted
against $\mu R$, where $R$ is the radius of $S^3$ (for comparison $0\leq\mu
L_s\leq6$ in Fig.~\ref{fig:muscan}). The action of QCD on $S^3
\times S^1$ to 1-loop order is given as in \cite{Hands:2010zp}, and was
originally derived in \cite{Aharony:2003sx} for more general matter content. At
low temperatures the partition function is given by 
\EQ{ \begin{aligned} {\cal Z}
(RT) &= \int \left[ {\mathrm d} \theta \right] e^{- S}\\ &= \int \left[
{\mathrm d} \theta \right] \text{exp}\left[ - \sum_{n=1}^{\infty} \frac{1}{n}
\left[ \Tr_A (P^n) + (-1)^n N_f z_f (\frac{n}{T R}, m R) e^{n \mu / T} \Tr_F
(P^n) \right] \right] , \end{aligned} } 
where $[d \theta] = \prod_{i=1}^{N} d
\theta_i$, $P = {\rm diag} \{ e^{i \theta_1}, e^{i \theta_2}, ..., e^{i
\theta_N} \}$, and $z_f (\frac{n}{T R}, m R)$ is given by
\SP{ z_f \left(\frac{n}{T R},mR
\right)&=\sum_{\ell=1}^\infty d_\ell^{(f)}e^{-n\varepsilon_\ell^{(f,m)}/T}\\
&=2\sum_{\ell=1}^\infty\ell(\ell+1)e^{-\frac{n}{T R}
\sqrt{(\ell+\frac12)^2+m^2R^2}}\ ,  } 
where $m$ is the quark mass. 
In Figure \ref{all_sphere}, the Polyakov
line ${\mathscr P}$, quark number ${\mathscr N}$, and quark number
susceptibility $\chi$, are all derived from the partition function. These are
given by 
\begin{eqnarray} \text{Polyakov line:} & \hspace{4mm} \PP &=
\frac{1}{\cal Z} \int \left[ {\mathrm d} \theta \right] e^{-S} \left( \sum_{i=1}^{N}
e^{i \theta_i} \right),\\ 
\text{Quark number:} & \hspace{4mm} {\mathscr N} &= T
\left( \frac{\partial \ln {\cal Z}}{\partial \mu} \right),\\ 
\text{Quark number
susceptibility:} & \hspace{4mm} {\chi} &= T \left( \frac{\partial \mathscr
N}{\partial \mu} \right).  \end{eqnarray} 
Note that with this normalisation $\chi$ is an extensive quantity.
Qualitatively the lattice results of
Figure \ref{fig:muscan} and the perturbative results on $S^3 \times S^1$ in
Figure \ref{all_sphere} appear similar: the fermion number rises in discrete
levels, and there is a spike in the Polyakov line and the quark number
susceptibility at each level transition. However, there are noticeable
differences in the shape of the curves during the transitions. The differences
between the lattice and perturbative results could be due to working at
different interaction strengths. Also, in small volumes effects from considering
different manifolds should be more apparent. In what follows we discuss each of
the observables calculated at $\beta = 24$ on the lattice in detail, comparing
with the results from $S^3 \times S^1$ where appropriate.

\begin{figure}[t]
\center
\includegraphics[width=13cm]{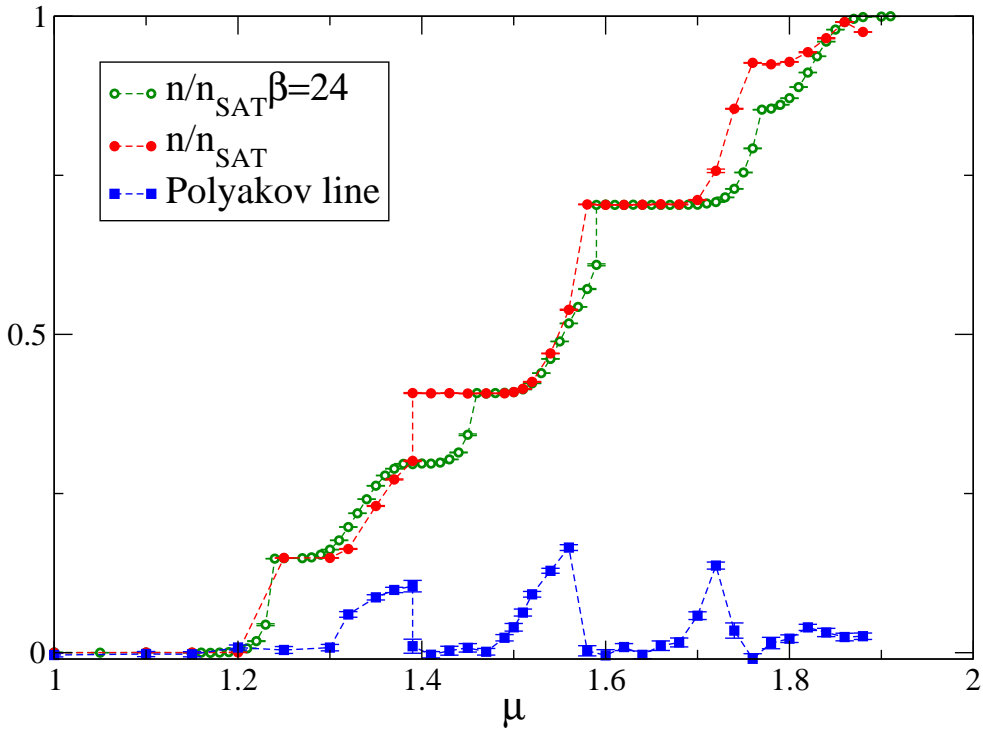}
\caption{Plot of $n_q/n_q^{\rm SAT}$ and ${\mathscr P}$ versus $\mu$ for $\beta=6$,
$\kappa=0.124$ on $3^3\times64$. Results for $\beta=24$ are also shown as open
symbols}
\label{fig:muscan_6.0}
\end{figure}

Before discussing each observable in detail it is interesting to see what happens when we increase the interaction strength a little more. Fig.~\ref{fig:muscan_6.0} shows the same scan of the $\mu$-axis at stronger
gauge coupling $\beta=6$. The level degeneracies corresponding to the
occupancies on the staircase are identical to those at $\beta=24$
indicating that the system responds adiabatically in this coupling range.
Moreover, the shell energies given by the risers are also little changed,
showing that dependence of the levels on $\beta$ is weak. The behaviour of
${\mathscr P}$
also suggests a similar relation with $n_q(\mu)$, albeit with a much
reduced signal which can be attributed to much larger quantum fluctuations at
this stronger coupling.

The most striking feature of Figs.~\ref{fig:muscan},\ref{fig:muscan_6.0} is the qualitative similarity
of the relation between $n_q$ and ${\mathscr P}$ to that found using weak coupling
methods on the hyperspherical attoworld of Figure \ref{all_sphere} and of the results in \cite{Hands:2010zp}. The numerical results reinforce the claim that
deconfining behaviour is correlated with a partially filled energy level or ``shell'' in
the box, implying that deconfinement is associated with a non-zero density of
gapless states.

\subsection{Quark number density $n_q$}

The weak-coupling results for $n_q(\mu)$ in Figure \ref{fig:muscan} exhibit a staircase
structure similar to that of Fig.~\ref{fig:free}, but with two important
differences. Firstly, the onset value of $\mu\sim1.2$ is considerably larger,
and secondly the degeneracies of the discrete levels are now 8, 8, 6, 16, 8 and
8, with the understanding that each of these numbers should be multiplied by a
factor $2N_f$ (see next paragraph). It should also be noted that over certain $\mu$-ranges $n_q$ is
double-valued; this arose from simulations where two distinct and apparently
stable, or at least metastable, states of the system were found (in such cases
the state with larger $n_q$ is shown with open symbols). In some cases
``tunnelling'' events from one state to the other occurred as the simulation
proceeded. Since both $L_s$
and $N_c$ are small, and we are accordingly far from any thermodynamic limit,
there seems to be 
no obvious  criterion for deciding which if either is the ``true'' ground
state.

Before commenting further on the level degeneracies, we wish to highlight an
important constraint on the interacting theory. The grand canonical partition
function can be written
\begin{equation}
{\cal Z}(\mu)=\sum_{\mathscr N} Z({\mathscr N})\exp({\mu\over T}{\mathscr N}),
\label{eq:canonical}
\end{equation}
where $Z$ is the canonical partition function evaluated in the sector with fixed
particle number ${\mathscr N}=n_qV$. On one of the steps in Fig.~\ref{fig:muscan} where
$n_q(\mu)$ is approximately constant, it must be the case that ${\cal Z}\propto\exp({\mu\over
T}{\mathscr N})$, implying that the sum in (\ref{eq:canonical}) is saturated by
one particular value of ${\mathscr N}$, and hence that ${\cal Z}\propto Z({\mathscr N})$. Now, using
the Fourier representation of the $\delta$-function,
(\ref{eq:canonical}) may be inverted to yield
\begin{equation}
Z({\mathscr N})={1\over{2\pi}}\int_{-\pi}^\pi d\vartheta e^{-i{\mathscr N}\vartheta}{\cal
Z}(i\vartheta T),
\end{equation}
{\it i.e.\/}~the canonical partition function is the ${\mathscr N}$th Fourier component of the grand
canonical partition function evaluated with imaginary chemical potential
$\mu_I=\vartheta T$.
However, it is known \cite{RW} that the $Z_{N_c}$ centre symmetry of the pure gauge theory can be
extended to the case of matter with imaginary chemical potential, resulting in a
periodicity 
\begin{equation}
{\cal Z}(i\vartheta T)={\cal Z}(iT(\vartheta+{{2\pi}\over N_c})).
\label{eq:RW}
\end{equation}
Eqns. (\ref{eq:canonical},\ref{eq:RW}) together imply $Z({\mathscr N})$ is only
non-vanishing for ${\mathscr N}/N_c$ integer, or in other words that the canonical
partition function is only defined for sectors of zero $N_c$-ality. 
Notice that, while the weak coupling analysis is obviously consistent with this constraint, the values of ${\mathscr N}$ observed on the steps of Fig.~\ref{fig:muscan} are, indeed,
multiples of $N_c$ which is a far from trivial check on the lattice analysis.

In fact, the observed pattern of degeneracies 8, 8, 6, 16, 8, 8  appears impossible to
explain in terms of single-particle plane wave states, in important contrast to
the free case Fig.~\ref{fig:free}. We conclude that while the staircase
behaviour of $n_q(\mu)$ appears to be a universal feature of attosystems,
non-perturbative effects continue to play an important role in determining the
spectrum.

\begin{figure}[t]
\center
\includegraphics[width=8cm]{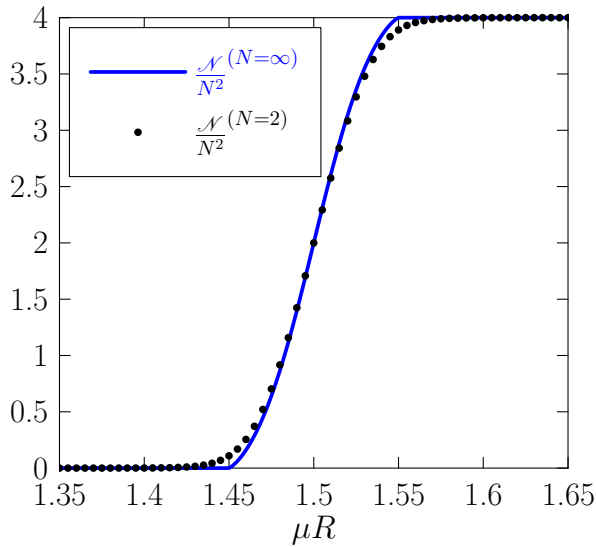}
\caption{Normalized quark number, ${\mathscr N} / N^2$, on $S^3 \times S^1$ for
the first level transition ($N_c = 2, \infty$). $N=N_f = N_c$, $m = 0$, $\frac{1}{T R} = 50$.}
\label{nferm_sphere}
\end{figure}

On $S^3 \times S^1$ the quark number at low temperature and zero quark mass is
given by \cite{Hands:2010zp} 
\EQ{ \begin{aligned} {\mathscr N} &= T \left(
\frac{\partial \ln{\cal Z}}{\partial \mu} \right)\\ &\xrightarrow[T \rightarrow 0]{}
\frac{N_f}{\cal Z} \int \left[ {\mathrm d} \theta \right] e^{-S}
\sum_{\ell=1}^{\infty} \sum_{i=1}^{N} 2 \ell (\ell+1) \left[
\frac{e^{\mu/T}}{e^{\mu/T} + e^{-i \theta_i + \frac{1}{T R} (\ell+\frac12)}}
\right] .  \end{aligned} } 
Here the level structure arises because of the
contribution from the Fermi-Dirac distribution function 
\EQ{ f(\varepsilon_\ell)
= \frac{1}{1+e^{(\varepsilon_\ell - \mu)/T}}, } 
which, at low temperatures, is
$1$ when $\mu$ is larger than an energy level $\varepsilon_\ell =
(\ell+{1\over2})/R$, and
zero otherwise. Thus the number of quarks at each level $L$ is a sum over the
fermion degeneracies $2\ell(\ell+1) N_c N_f$ of all levels with $\varepsilon_\ell <
\mu$: 
\EQ{ {\mathscr N}_L = N_c N_f \sum_{\ell=1}^{L} 2 \ell (\ell+1)\ .  
\label{eq:N_L}} 
A
plot of the quark number on $S^3 \times S^1$ as a function of $\mu R$ for $N_c =
2$, and $N_c = \infty$, is given in Figure \ref{nferm_sphere}. In both cases the
quark number is symmetric about the mid-point of the transition at
$\mu=\varepsilon_\ell$, in that the level transition starts and finishes at
the same rate, in contrast to the lattice results of Fig.~\ref{fig:muscan}. 
The $N_c = \infty$ results also shown
(from \cite{Hands:2010zp}), exhibit quite clearly the two Gross-Witten-type phase  
transitions as the points of non-analyticity where the quark number first turns on and then saturates as $\mu R$ is increased.

\subsection{Polyakov line ${\mathscr P}$}

In addition to quark density $n_q$, the other important observable monitored was
the Polyakov line defined in Eqn. (\ref{eq:L}).  In theories without fundamental
matter ${\mathscr P}$ is an order parameter for deconfinement; even in the
presence of quarks it can be related to the free energy $f_Q$ of a static fundamental
source via ${\mathscr P}\sim\exp(-f_Q/T)$, and lattice studies of eg.  the thermal
transition in QCD observe a sharp increase in ${\mathscr P}$ around the
deconfining temperature. The behaviour of ${\mathscr P}$ in systems with
$\mu\not=0$ and $L_t>L_s$ is considerably less studied, although intriguing
results have been reported in \cite{Hands:2006ve,Hands:2010gd}. Note that for
$N_c=2$ there is no distinction  between ${\mathscr P}$ and ${\mathscr P}_{-1}$
defined using the inverse link variables $U^\dagger$ in (\ref{eq:L}).

Fig.~\ref{fig:muscan} shows that the Polyakov line ${\mathscr P}(\mu)$ exhibits a
complicated
behaviour: in regions where $n_q(\mu)$ is constant its value is small, possibly
consistent with zero, but in regions where $n_q$ is changing ${\mathscr P}>0$, implying
deconfinement. In fact it exhibits rather sharp maxima over at least six
distinct ranges of $\mu$: its numerical value ${\mathscr P}_{max}\simeq0.4$ is in accord
with expectations that the theoretical maximum value of 1 is renormalised downwards due to quantum
fluctuations~\cite{Gupta:2007ax}.
When $n_q$ increases to one of its plateaux over a very short $\mu$-interval, the transition from 
${\mathscr P}>0$ to ${\mathscr P}\approx0$ is very sudden; on other occasions when the $n_q(\mu)$ behaviour
between plateaux is an extended S-shape, ${\mathscr P}(\mu)$ has a more 
symmetrical peak. This seems to be true even in regions where both
confining and deconfining states are found.
\begin{figure}[t]
\center
\includegraphics[width=13cm]{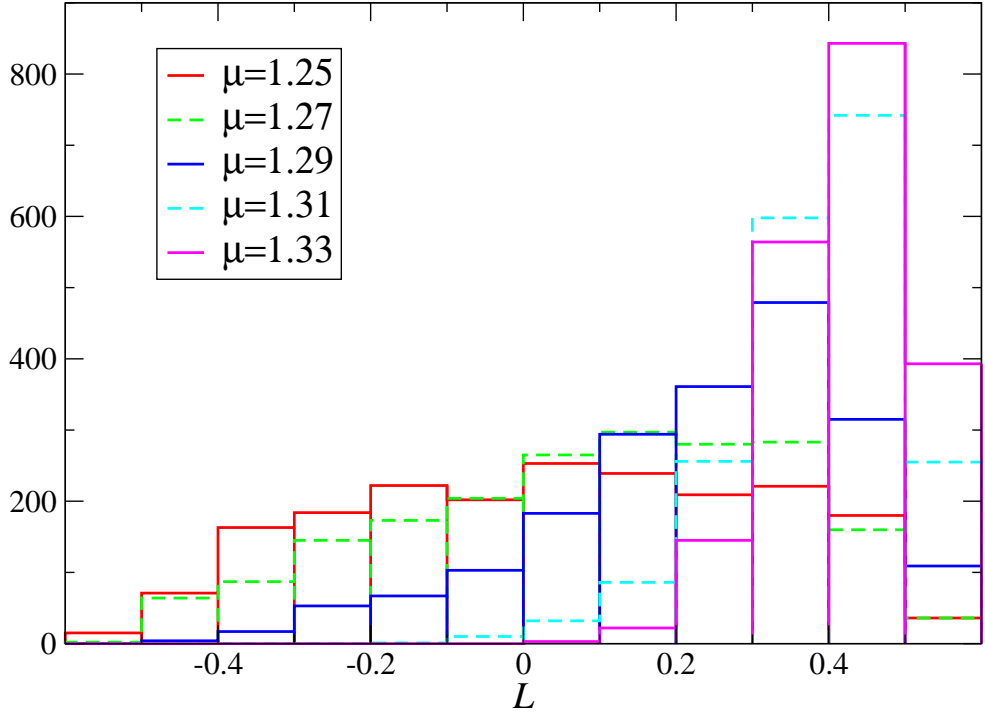}
\caption{Histogram of ${\mathscr P}$ over a sample of
2000 configurations with $\beta=24$ at various $\mu$.}
\label{fig:polyhist}
\end{figure}
Fig.~\ref{fig:polyhist} focusses on the range $1.23\leq\mu\leq1.33$ over
which the second shell is gradually occupied and $L$ rises from zero to its
maximum value. The histogram shows that the fluctuations $\delta{\mathscr P} /{\mathscr
P}_{max}\sim
O(1)$ throughout, but that the distribution evolves from being symmetrically
centred on zero at $\mu=1.25$ to being highly skewed at $\mu=1.33$.

\begin{figure}[t]
  \hfill
  \begin{minipage}[t]{.49\textwidth}
    \begin{center}
\includegraphics[width=0.99\textwidth]{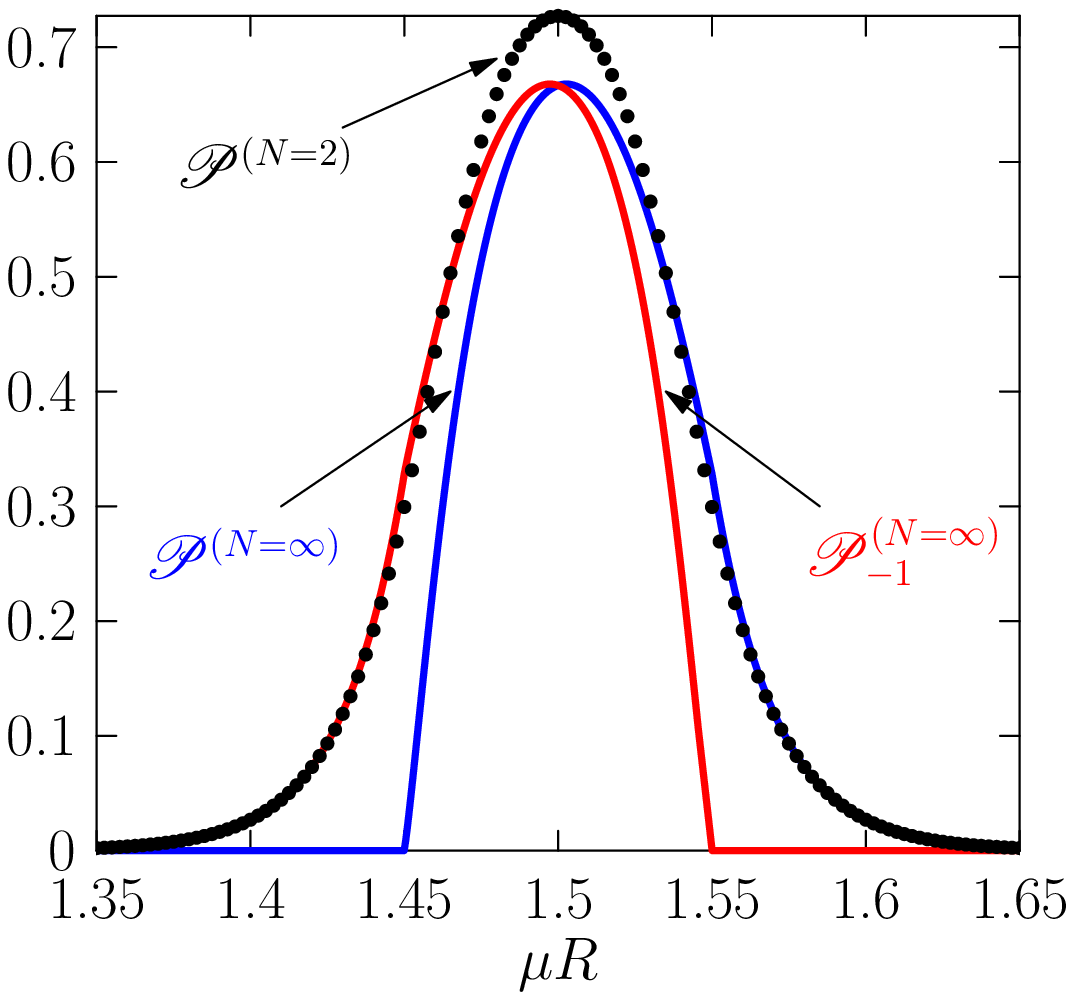}
    \end{center}
  \end{minipage}
  \hfill
  \begin{minipage}[t]{.49\textwidth}
    \begin{center}
\includegraphics[width=0.99\textwidth]{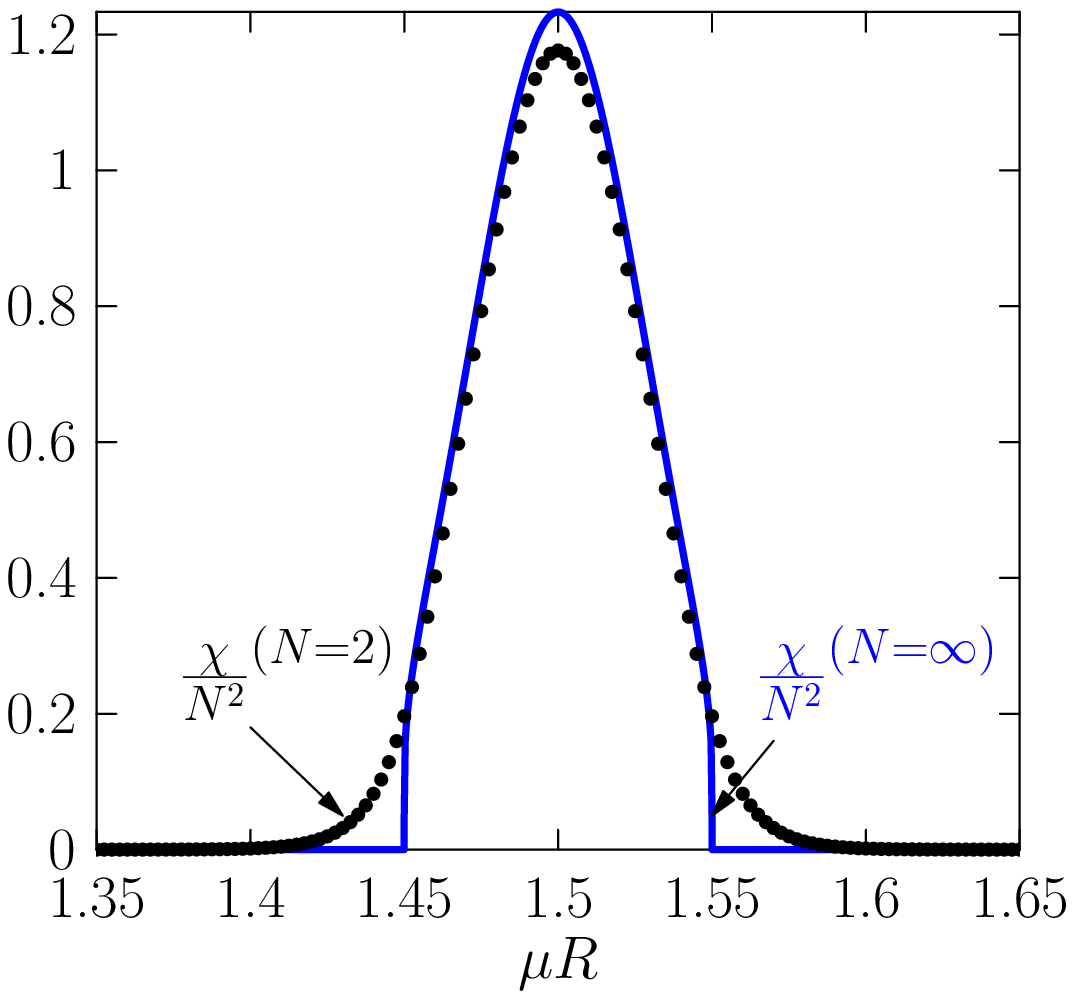}
    \end{center}
  \end{minipage}
  \hfill
\caption{Polyakov lines (left) and normalized quark number susceptibility (right) 
on $S^3 \times S^1$ for the first level transition ($N_c = 2, \infty$). 
$N=N_f = N_c$, $m = 0$, $\frac{1}{T R} = 50$.} 
\label{poly_susc_sphere}
\end{figure}

The Polyakov lines on $S^3 \times S^1$ as a function of $\mu R$ are shown in
Fig.~\ref{poly_susc_sphere} (left) for $N_c = 2$ and $N_c = \infty$. For $N_c
= 2$, ${\mathscr P} = {\mathscr P}_{-1}$ and the Polyakov line is symmetrical
around each energy level $\mu = \varepsilon_\ell$. It is important to note that
this is
not the case for $N_c \ne 2$, where ${\mathscr P} \ne {\mathscr P}_{-1}$ and
these are individually asymmetrical around each energy level, though mirror
images
of each other. This reflects that fact that in general fundamental
and antifundamental test charges respond differently to a background density
of fundamental charge.  As remarked above, the lattice results 
show both features which are approximately symmetrical as well as some which
are clearly skewed, in apparent contradiction with the $N_c=2$ prediction shown
in Fig.~\ref{poly_susc_sphere}.

\subsection{Quark number susceptibility $\chi_q$}

To a very good approximation Fig.~\ref{fig:muscan} shows that 
${\mathscr P}\propto\chi_q$;
indeed, the quark number susceptibility is
often taken as an alternative signal of deconfinement in lattice studies of
the thermal QCD transition~\cite{Bazavov:2009zn}. 

Figure \ref{poly_susc_sphere} (right) shows the quark number susceptibility from perturbation theory on $S^3 \times S^1$ for $N_c = 2$ and $N_c = \infty$. For $N_c = 2$ the quark number susceptibility is calculated by numerical integration:
\begin{equation}
\begin{aligned}
\chi &= T \frac{\partial {\mathscr N}}{\partial \mu}
= T^2 \frac{\partial}{\partial \mu} \left[ \frac{1}{\cal Z} 
\int [{\rm d} \theta] e^{-S} \left( \frac{\partial S}{\partial \mu} \right) \right]\\
&= - T^2 \left[ \frac{1}{{\cal Z}^2} \left( \int [{\rm d}\theta] e^{-S} 
\left( \frac{\partial S}{\partial \mu} \right) \right)^2 - 
\frac{1}{\cal Z} \int [{\rm d}\theta] e^{-S} \left( \frac{\partial S}{\partial \mu} \right)^2 + 
\frac{1}{\cal Z} \int [{\rm d}\theta] e^{-S} \left( \frac{\partial^2 S}{\partial \mu^2} \right) \right] ,
\end{aligned}
\end{equation}
where
\begin{equation}
\frac{\partial S}{\partial \mu} = - \frac{N_f}{T} \sum_{i=1}^{N_c}
\sum_{\ell=1}^{\infty} 2\ell(\ell+1)\left[
\frac{1}{1+e^{\frac{1}{T}(\varepsilon_\ell - \mu) - i \theta_i}} \right]
\end{equation}
and
\begin{equation}
\frac{\partial^2 S}{\partial \mu^2} = - \frac{N_f}{T^2} \sum_{i=1}^{N_c}
\sum_{\ell=1}^{\infty} 2\ell(\ell+1)\left[ \frac{e^{\frac{1}{T}
(\varepsilon_\ell - \mu) - i \theta_i}}
{(1 + e^{\frac{1}{T} (\varepsilon_\ell - \mu) - i \theta_i})^2} \right] .
\end{equation}

To obtain the $N_c = \infty$ result the quark number susceptibility is calculated by taking 
the derivative of the result for the quark number in \cite{Hands:2010zp}. 
It is only non-zero while a level transition is taking place; near the $L^{th}$
step the expression is
\begin{equation}
\frac{\chi}{N_c^2} = T \frac{\partial {\cal N}}{\partial \mu} = \biggl(\ln \left[ \frac{(1+{\cal N})
(1+{\mathscr N}_L - {\cal N})}{({\mathscr N}_L - {\cal N}) {\cal N}} \right]\biggr)^{-1}
\label{susc_Ninf}
\end{equation}
where ${\cal N} (\mu)$ is determined as in \cite{Hands:2010zp} from numerical inversion of
\begin{equation}
\xi_\ell \equiv e^{\frac{1}{T} (\mu - \varepsilon_\ell)} = 
\frac{({\mathscr N}_L - {\cal N})^{{\mathscr N}_L - {\cal N}} 
(1 + {\cal N})^{1+{\cal N}}}{{\cal N}^{\cal N} (1 + {\mathscr N}_L- {\cal N})^{1
+ {\mathscr N}_L- {\cal N}}} ,
\end{equation}
and ${\mathscr N}_L$ is given by (\ref{eq:N_L}). 
It is clear from eq. (\ref{susc_Ninf}) and Figure \ref{poly_susc_sphere} (right)
that for $N_c = \infty$ another derivative of $\chi_q$ with respect to $\mu$ 
will result in discontinuities at the beginning and end of
each level transition, indicating that these correspond to third order
transitions as predicted in \cite{Hands:2010zp}. 

We note that the level structure of the particle number and
corresponding susceptibility spikes were also observed in
simulations of the non-linear sigma model \cite{Banerjee:2010kc}.

\subsection{Quark--antiquark condensate $\langle\bar\psi\psi\rangle$}

\begin{figure}[t]
\center
\includegraphics[width=13cm]{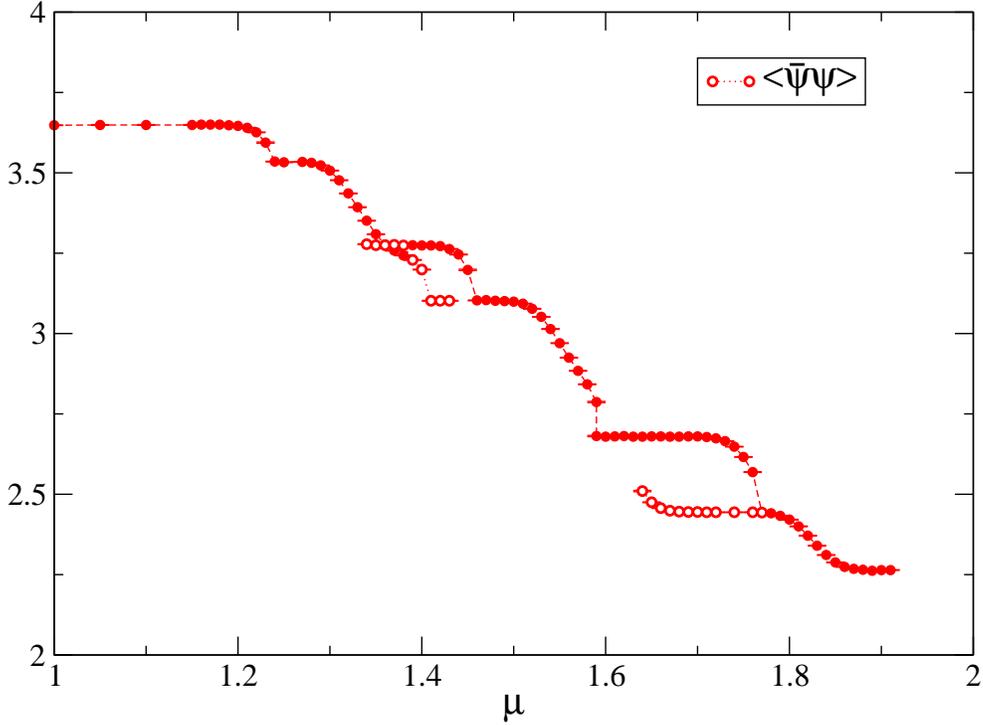}
\caption{Plot of $\langle\bar\psi\psi\rangle/4$ versus $\mu$ for $\beta=24$,
$\kappa=0.124$ on $3^3\times64$. Open symbols have the same meaning as
Fig.~\protect \ref{fig:muscan}.}
\label{fig:psibarpsi}
\end{figure}
Next we discuss another fermionic observable, the quark--antiquark condensate,
which requires a non-perturbative calculation. For Wilson lattice fermions
(\ref{eq:M}):
\begin{equation}
\langle\bar\psi\psi\rangle=4N_cN_f-\kappa{T\over V}{{\partial\ln{\cal
Z}}\over{\partial\kappa}}.
\end{equation}
Because (\ref{eq:M}) has no chiral symmetry, this bilinear has no
interpretation as an order parameter, but does yield information on the
conformal anomaly~\cite{Hands:2006ve}. The data, however, echo the structure of
Fig.~\ref{fig:muscan}: $\langle\bar\psi\psi\rangle$ is approximately constant in
$\mu$-ranges where there are only filled shells, and changes rapidly in the same
regions where $n_q$ changes rapidly, corresponding to a partially-filled shell.

On $S^3 \times S^1$ the quark-antiquark 
condensate is calculated using \cite{Hands:2010zp}
\SP{
\langle {\bar \psi} \psi \rangle &= - \frac{T}{V_3} 
\left( \frac{\partial \ln{\cal Z}}{\partial m} \right)\\
&\xrightarrow[T \rightarrow 0]{} \frac{N_f
  m}{\pi^2 R^2} \int \left[ {\mathrm d} \theta \right] e^{-S}
\sum_{\ell=1}^{\infty} \sum_{i=1}^{N} \frac{\ell
  (\ell+1)}{(\ell+\frac{1}{2})} \left[ \frac{e^{\mu/T}}{e^{\mu/T} + e^{-i
\theta_i + \frac{!}{T} \varepsilon^{(f,m)}_\ell}} \right]\ ,
}
which gives zero when $m = 0$,  and a level structure qualitatively similar to
Fig.~\ref{fig:psibarpsi} if $m\not=0$. 
\subsection{Gluonic observables}

\begin{figure}[t]
  \hfill
  \begin{minipage}[t]{.49\textwidth}
    \begin{center}
\includegraphics[width=0.99\textwidth]{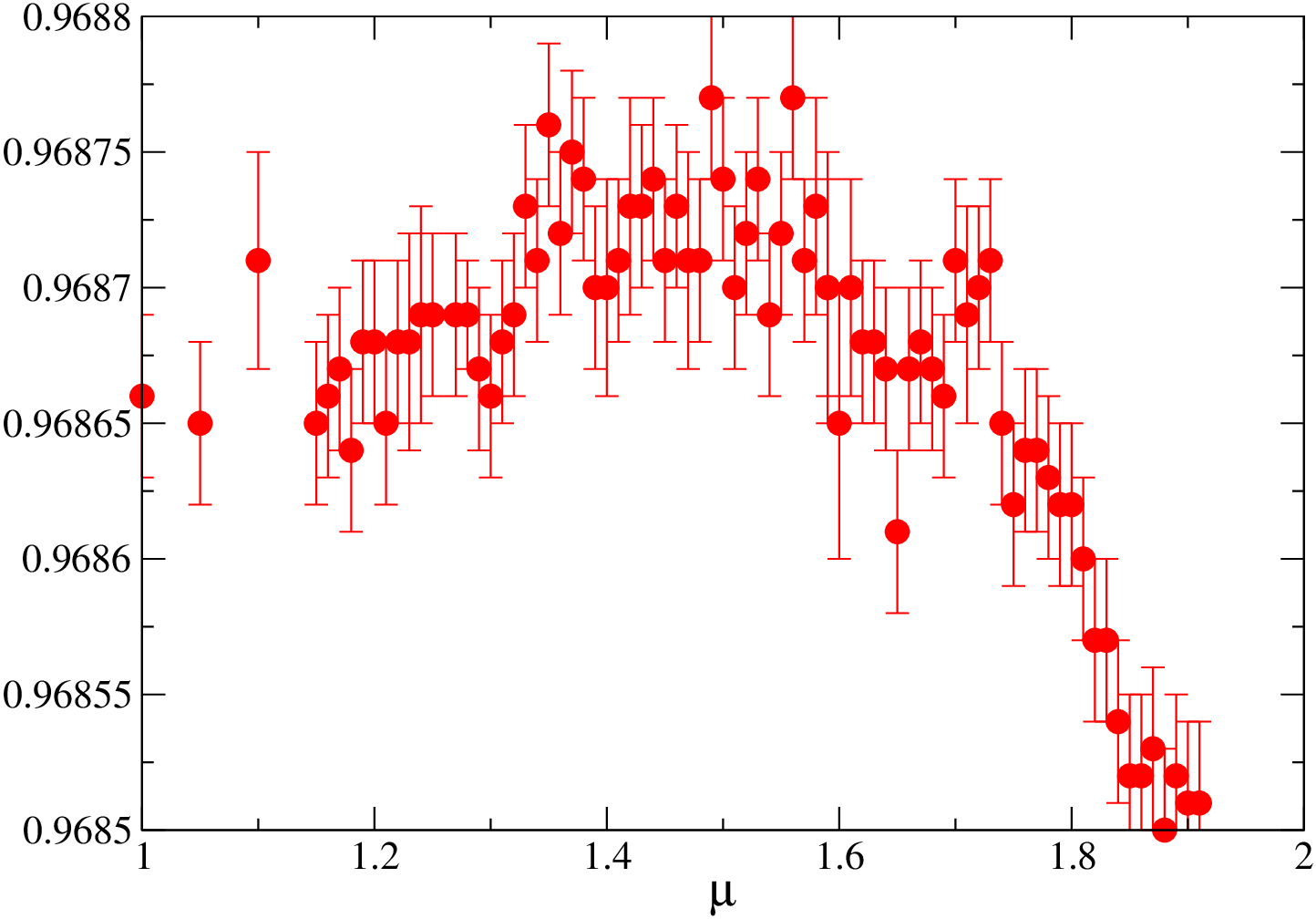}
    \end{center}
  \end{minipage}
  \hfill
  \begin{minipage}[t]{.49\textwidth}
    \begin{center}
\includegraphics[width=0.99\textwidth]{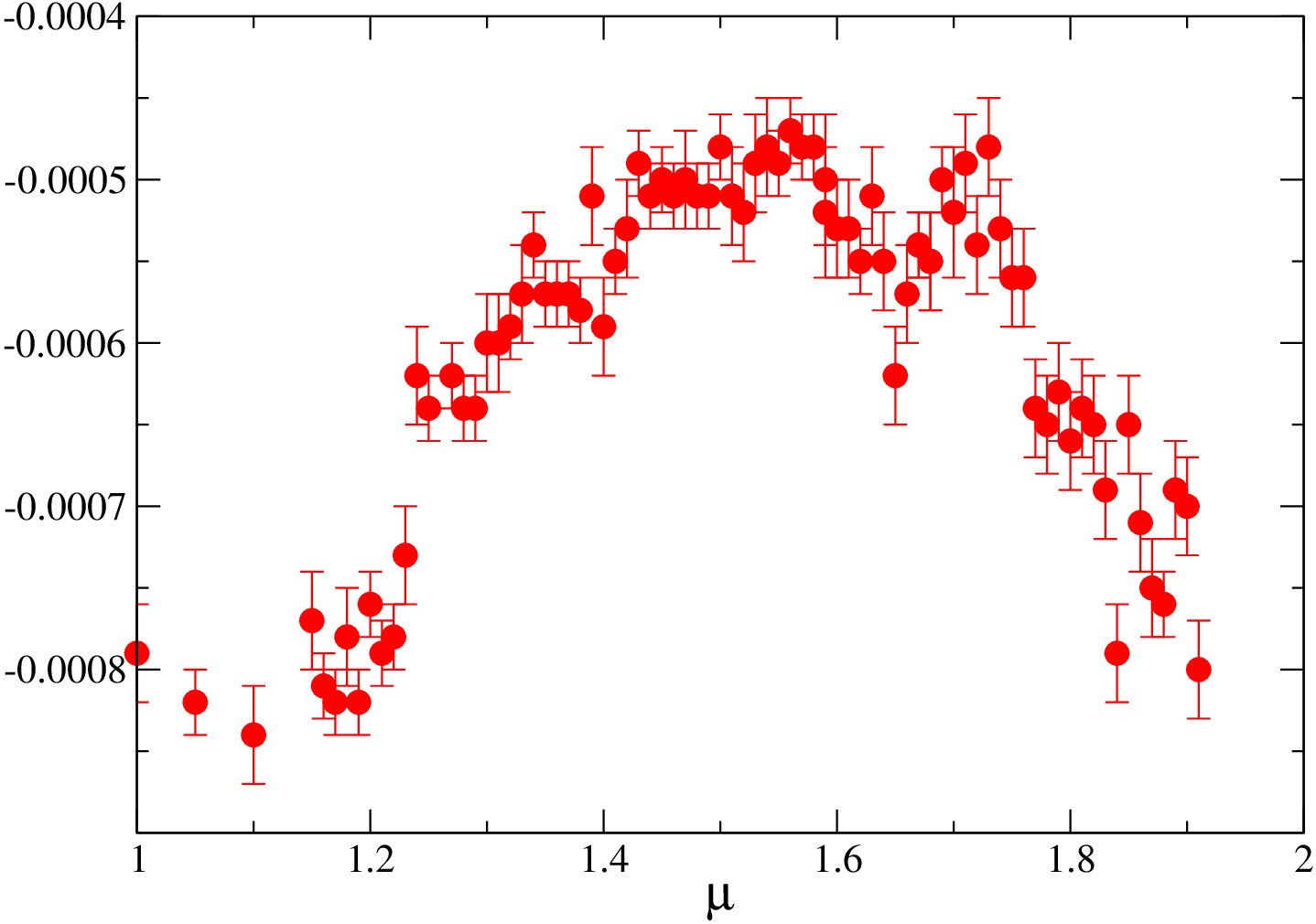}
    \end{center}
  \end{minipage}
  \hfill
\caption{Gluon observables as a function of $\mu$ for $\beta=24$.
(Left): ${1\over2}(\Box_t+\Box_s)$ (Right): $\Box_t-\Box_s$.}
\label{fig:gluons}
\end{figure}
It is also of interest to consider gluonic observables. The simplest local
gauge-invariant gluon observable is the plaquette $U_{\nu\lambda}$. In a
non-Lorentz invariant system such as one with $\mu\not=0$ it is helpful to
define
\begin{equation}
\Box_s={1\over3N_c}{T\over
V}\sum_x\sum_{i<j}\langle\mbox{tr}U_{ijx}\rangle;\;\;\
\Box_t={1\over3N_c}{T\over
V}\sum_x\sum_{i}\langle\mbox{tr}U_{0ix}\rangle.
\end{equation}
The normalisations ensure $\Box_{s,t}\to1$ in the $\beta\to\infty$ limit. We 
then consider in Fig.~\ref{fig:gluons} both the average plaquette ${1\over2}(\Box_s+\Box_t)$ and the
difference $\Box_t-\Box_s$, which is proportional to the gluon energy density
\begin{equation}
\varepsilon_g=3Z\beta(\Box_t-\Box_s),
\end{equation}
where $Z$ is a renormalisation factor calculable in perturbation
theory~\cite{Karsch:1982ve}.

The average plaquette has a value extremely close to unity, as befits such a
weak coupling. It has a non-monotonic variation with $\mu$, qualitatively similar to
the behaviour found on larger systems in \cite{Hands:2006ve,Hands:2010gd}. The
decrease at large $\mu$ is readily understood as a consequence of saturation
$n_q/n_q^{\rm SAT}\to1$: in this regime screening due to virtual quark -- antiquark
pairs is suppressed due to Pauli blocking, and hence many gluonic observables
revert to their values in the quenched theory, which has in effect a larger
lattice spacing and hence a larger departure from the free-field value. 
At $\mu\approx1.7$, where the average plaquette dips below its value at
$\mu=0$, the ratio $n_q/n_q^{\rm SAT}\approx0.7$, to be contrasted with
the corresponding value $\sim0.018$ observed at $\beta=1.9$~\cite{Hands:2010gd}. 

The difference $\Box_t-\Box_s$ also shows non-monotonic behaviour; its negative
value makes it hard to interpret as a physical energy density, and probably
arises as
an artifact of the lattice aspect ratio  $L_t\gg L_s$. Both plots hint at a finer
structure, such as a dip at $\mu\approx1.65$. Curiously this does not appear to
match any interesting region of Fig.~\ref{fig:muscan}, except in the sense that
both confining and non-confining solutions appear to be stable here.

\subsection{Conjugate gradient iterations}

\begin{figure}[t]
\center
\includegraphics[width=13cm]{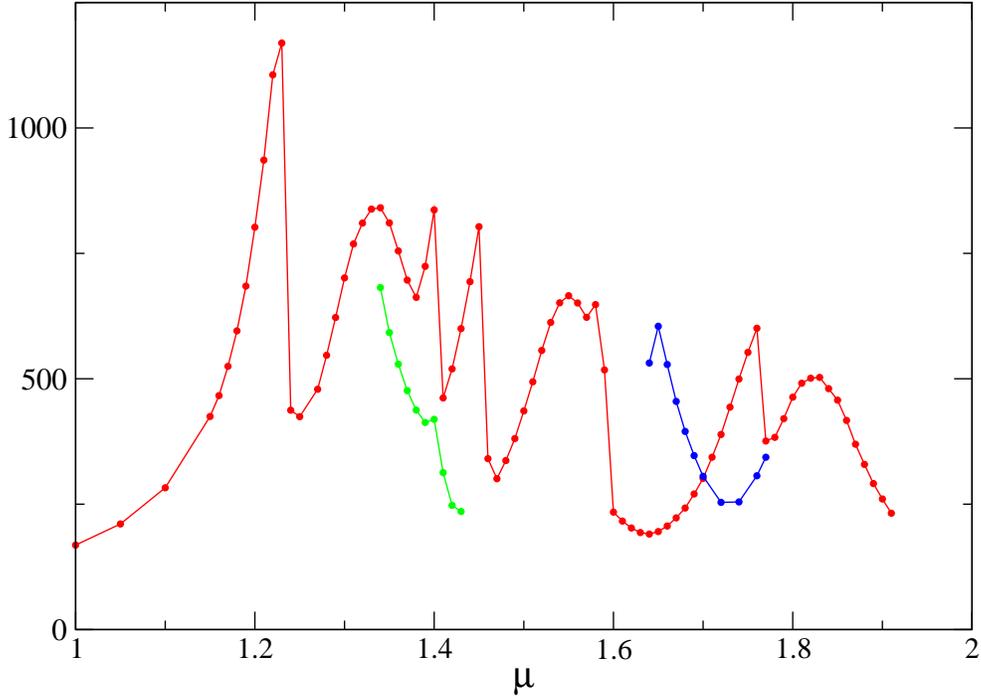}
\caption{Number of {\tt congrad} iterations required in the acceptance step of
the HMC algorithm versus $\mu$ for $\beta=24$, $\kappa=0.124$ on $3^3\times64$.
Blue and green points correspond to the open symbols of Fig.~\protect \ref{fig:muscan}.}
\label{fig:congrad}
\end{figure}
Finally, Fig.~\ref{fig:congrad} plots the number of iterations of the conjugate
gradient algorithm required to invert $M$ to some pre-specified accuracy (in
this case such that the average norm of the residual vector should not exceed
$10^{-9}$ per site, spin and color) as a function of $\mu$.
This quantity carries physical information since it is roughly inversely
proportional to $\vert\lambda_{min}\vert^2$, where $\lambda_{min}$ is the
complex eigenvalue of $M$ closest to the origin. Fig.~\ref{fig:congrad} suggests
that the smallest eigenvalue evolves smoothly so long as $n_q(\mu)$ is smooth,
but jumps sharply at the same points where $n_q(\mu)$ is discontinuous. 
Note that the eigenvalue is always smallest in
deconfined regions where $\chi_q, {\mathscr P}>0$, again consistent with the picture of a non-zero
density of gapless modes.
Interestingly, this pattern is also
respected by the alternative solutions corresponding to the open symbols of
Fig.~\ref{fig:muscan}, suggesting that at these points the simulation sometimes
evolves smoothly with $\mu$, and at other times
jumps to a new ground state with significantly smaller $\lambda_{min}$;
further studies on systems with varying $L_t$ (correpsonding to varying $T$)
might clarify the situation here.

\section{Summary}
\label{sec:discussion}

This paper has followed up the study of QCD at non-zero chemical potential on
spaces defined by a scale $R\ll\Lambda_{QCD}^{-1}$ initiated in
Ref.~\cite{Hands:2010zp} by presenting Monte Carlo results for the case
$N_c=2$
which is simulable using orthodox lattice gauge theory techniques. We have shown
that the two principal qualitative features of the perturbative calculation on
the hypersphere, namely that the quark density $n_q$ is a step-wise function of
$\mu$, and that the Polyakov line ${\mathscr P}$ is significantly different from
zero only in the $\mu$-ranges where $n_q$ is rising, persist on the three-torus
$(S^1)^3\times S^1$ even once non-perturbative effects are correctly included.
The staircase form of $n_q(\mu)$ is indicative of a series of widely separated
sets of physical states akin to the shell structure of, say, a nucleus.
Moreover we have identified a further relation, ${\mathscr
P}\propto\chi_q={{\partial n_q}/{\partial\mu}}$,
which seems to be respected equally well in both approaches.
These results lend support to the interpretation of $\chi_q$ as an alternative
indicator of confinement/deconfinement in systems with fundamental matter, and
hint at a relation between deconfinement and a non-vanishing density of gapless
states (in other words, the existence of a ``conduction band'').

However, the two approaches differ in their detailed predictions for the
shell degeneracies. Comparison of simulation results at two different $\beta$
values suggests the system evolves adiabatically with coupling, but it has not
proved possible to interpret the levels in terms of the single-quark states
underpinning the perturbative approach, and a
full explanation of the structure revealed in Fig.~\ref{fig:muscan} is still
missing.

In addition, identification of the correct ground state of the system as $\mu$
is varied was troublesome due to the apparent existence of more than one
``solution'' stable under HMC evolution (in fact, this proved to be a much more
serious problem in pilot studies on $2^3\times64$, and persuaded us to switch to
a system where $L_s$ was not a multiple of $N_c$).
In conventional simulation campaigns
such ambiguities are normally resolved by taking the thermodynamic limit, 
but here $L_s$ and $N_c$ are both finite. It would, of course, be interesting
to try a simultaneous tuning of $L_s$ and $\beta$, keeping
$L_sa$ fixed, but this requires resources considerably beyond what we have
been able to expend. In any case, more  general questions about how both thermodynamic 
and zero temperature limits are approached would also be interesting to explore.

\section{Acknowledgements}
This project was enabled with the assistance of IBM Deep Computing.

\clearpage
\end{document}